# Switchable circular dichroism and ionic migration dominated charge transport in a chiral spin crossover polymer

M. Zaid Zaz,[1] Sartaz Sakib,[2] Wai Kiat Chin,[1] Peace Adegbite,[1] Gauthami Viswan,[1] Alpha T'Ndaiye,[3] Andrew J Yost,[1] Rebecca Y Lai,[2] and Peter A Dowben[1]

[1]Department of Physics and Astronomy, Jorgensen Hall, University of Nebraska-Lincoln, Lincoln, NE 68588-0299, United States of America

[2]Department of Chemistry, Hamilton Hall, University of Nebraska-Lincoln, Lincoln, NE 68588-0304, United States of America

[3]Advanced Light Source, Lawrence Berkeley National Laboratory, Berkeley, CA 94720, United States of America

**Abstract**
We demonstrate thermally switchable chiroptical activity in a chiral spin-crossover (SCO) material, where circular dichroism is significant in the low spin state but quenched in the high spin state for both enantiomers. With magnetometry we establish a cooperative transition with hysteresis near room temperature. Fe $L_{3,2}$-edge X-ray absorption directly links the quenching of chirality to a reorganization of Fe 3d electronic structure that accompanies the spin state transition. Moreover, electrical measurements show pronounced I(V) hysteresis and cycling dependent C(V) behavior, indicating transport in this chiral SCO material is dominated by ionic migration rather than mobile electrons or holes.

**Introduction**
Spin crossover (SCO) compounds are among the most compelling molecular switching platforms because a single coordination framework can be reversibly driven between low-spin and high-spin electronic configurations, often with cooperative thermal hysteresis that enables bistable states [1-3]. In chiral SCO systems, which have been widely studied [4-26] this electronic reconfiguration can, in principle, reshape the electron density landscape and therefore modulate chirality [26]. Building on this idea, we investigate the chiral molecular system [Fe(NH$_2$trz)$_3$](X-CSA), (where trz= triazole, CSA= camphorsulphonate, and X= L/D) whose circular dichroism can be tuned by driving it across the spin crossover transition, and we demonstrate that the chiroptical response is effectively switched between the low-spin and high-spin states. While the [Fe(NH$_2$trz)$_3$](L-CSA) analogue has previously been synthesized and shown to exhibit spin-state-dependent chirality, we report the successful synthesis and the observance of spin state dependent circular dichroism in the [Fe(NH$_2$trz)$_3$](D-CSA) enantiomer, thereby completing the enantiomeric pair within the same molecular framework [26]. We establish the spin transition using temperature-dependent magnetic susceptibility, then show directly by UV–vis circular dichroism that the low-spin state exhibits appreciable dichroism while the high-spin state shows a strongly quenched response for both enantiomers.

A natural implication is that such a platform appears ideal for exploring chiral induced spin selectivity (CISS) [27-28], since one might expect that turning chirality "on" and "off" across the SCO transition could provide an equally direct handle on spin-selective transport [29]. However, the practical accessibility of CISS hinges not only on chirality, but also on whether charge transport proceeds in an electronically dominated regime where spin-dependent transmission through a chiral potential can be meaningfully exploited [30-34]. In this work, we therefore pair the chiroptical switching with a metal-centered electronic structure probe and an explicit transport diagnosis. Fe $L_{2,3}$-edge X-ray absorption spectroscopy, which has been extensively used to characterize spin crossover transitions and accompanying electronic redistribution [24, 35-40] is able to correlate the quenching of CD with a rearrangement of metal 3d unoccupied states across the transition. Thus, linking the loss of optical chirality to a concrete electronic reconfiguration at the Fe center. We then test the transport regime directly and find that the electrical response is dominated by ionic migration in a highly dielectric medium, as evidenced by pronounced I–V hysteresis and cycling-dependent C–V behavior with relaxation and recovery [30-34]. With this we re-emphasize that to achieve SCO-tunable chirality, switchable chirality is necessary, but if transport is governed by ionic motion and dielectric polarization, CISS signatures can be suppressed or rendered unusable even when chiroptical switching is robust.

**Materials and Methods**

*Synthesis*
The synthesis follows adaptations of the procedure reported in [26] . First, the precursor Fe(X-CSA)$_2$·6H$_2$O, (X=L,D) was prepared. 50 mg iron powder (0.9 mmol) was added in small portions to a solution containing 500 mg X-camphor sulfonic acid (2.16 mmol), (X=L,D) dissolved in 2 mL water. The mixture was stirred at 80 °C until the iron was no longer soluble. The resulting solution was then refluxed slowly at 130 °C for five hours and subsequently allowed to cool and crystallize overnight. Pale green crystals of Fe(X-CSA)$_2$·6H$_2$O formed, which were separated by filtration and dried in air then a solution containing 10 mg Fe(X-CSA)$_2$·6H$_2$O (0.016 mmol) in 125 µL DMSO and 2 mL acetonitrile was prepared by dissolving the compound first in DMSO and then adding acetonitrile. Separately, 10 mg of 4-amino-1,2,4-triazole (0.119 mmol) was dissolved in 125 µL DMSO and 2 mL acetonitrile in the same manner. The two solutions were mixed rapidly, producing a turbid pink colloid within a few minutes. The mixture was allowed to react for 12 hours in a closed container at room temperature and was then centrifuged for 15 minutes at 8000 rpm. The deposited pink solid was washed with acetonitrile. The final product [Fe(NH$_2$trz)$_3$](X-CSA), (X=L,D) was thus obtained.

*Magnetic susceptibility*
Molar magnetic susceptibility measurements were recorded on a VersaLab 3 Tesla Cryogen-Free Vibrating Sample Magnetometer (VSM) with an applied magnetic field of 1 T from 210-370K at sweeping rate 1K/min.

*Circular dichroism*
Circular dichroism spectra were recorded on a Jasco J815 CD spectrometer that was equipped with a thermos-stated cell holder. The temperature was changed at a rate of 2 K/min.

*X-ray absorption spectroscopy*
X-ray absorption spectroscopy measurements were recorded on the magnetic spectroscopy beamline 6.3.1 of the Advanced Light Source at the Lawrence Berkeley National Laboratories.

The spectra were taken at the Fe L$_{3,2}$ absorption edge in the absence of any applied magnetic field with an energy step of 0.1 eV.

*Transport measurements*
I(V) characteristics and small signal C versus V measurements were recorded using a lake shore cryogenic probe station with a Keithley 4200A-SCS Parameter Analyzer.

**Results and discussion**

*Magnetic susceptibility*
Temperature-dependent molar magnetic susceptibility measurements establish that both isomers of [Fe(NH$_2$trz)$_3$](X-CSA), (where trz= triazole, CSA= camphorsulphonate and X= L/D) undergo a thermally driven spin crossover between a low-spin (LS, diamagnetic) state and a high-spin (HS, paramagnetic) state in the 300–310 K range (Fig. 1). The transition for the L enantiomer occurs at a slightly lower temperature than for the D enantiomer, while the overall temperature span of the crossover is comparable for both the enantiomers. Both compounds exhibit a cooperative transition accompanied by a clear thermal hysteresis, and the hysteresis widths are similar in magnitude. This behavior is consistent with the established phenomenology of polymeric Fe(II) triazole spin-crossover systems and other spin crossover polymers [41-43]. The susceptibility data therefore provide the thermal set points used throughout this work to define LS and HS regimes for the optical, X-ray spectroscopic, and electrical measurements discussed below. In practical terms, the magnetic data demonstrate that the two counterion enantiomers retain comparable SCO cooperativity and that the handedness of the CSA counterion does not suppress the spin transition.

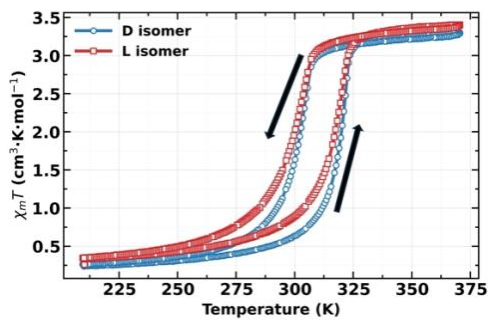

Fig 1. Molar magnetic susceptibility times temperature vs temperature
plot for [Fe(NH$_2$trz)$_3$](D-CSA) (blue) and [Fe(NH$_2$trz)$_3$](D-CSA) (red),
Indicating a temperature driven transition between a low spin (diamagnetic)
state and high spin (paramagnetic) state. The arrows indicate increasing and
decreasing temperature cycles.

*Circular dichroism*
UV–vis circular dichroism (CD) spectra measured for both enantiomers reveal substantial chiroptical activity across 220–400 nm together with a strong dependence on the thermally selected spin state (Fig. 2). In the LS regime, each compound exhibits a CD spectrum with well-defined features, indicating that the complex possesses an optically active excited-state manifold in this energy range. Upon heating into the HS regime, the CD magnitude is strongly reduced for both X = D and X = L, and the overall spectral contrast between LS and HS is readily resolved under the measurement conditions. Importantly, the temperature dependence of the CD response follows the same qualitative trend for the two enantiomers, namely a larger CD signal in the LS state and a diminished response in the HS state. This shared LS-to-HS suppression is consistent with chirality being encoded by the handed counterion environment and the chiral arrangement it imposes on the coordination framework, while the SCO transition modulates the magnitude of the CD through changes in the metal-centered electronic configuration and its coupling to ligand-based transitions. The observed reversibility within the measurement protocol indicates that the CD switching is not a result of irreversible sample degradation across the temperature range studied, but rather tracks a reversible thermally driven change in the underlying electronic structure. The CD data therefore show that both members of the enantiomeric pair display spin-state-dependent chiroptical contrast, providing a consistent optical readout of SCO in these complexes.

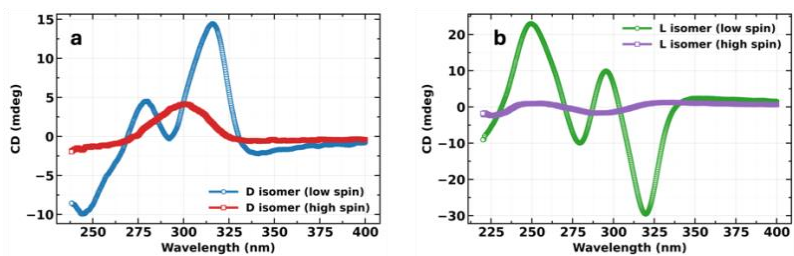

Fig 2. Circular dichroism vs wavelength plot for a. [Fe(NH$_2$trz)$_3$](D-CSA) in the high and low spin state and for b. [Fe(NH$_2$trz)$_3$](L-CSA) in the high and low spin state. The circular dichroism in the high spin state is significantly quenched for both the isomers.

X-ray absorption spectroscopy
To relate the observed chiroptical changes to an Fe-centered electronic marker of the spin state, Fe L$_{2,3}$-edge X-ray absorption spectra were acquired for [Fe(NH$_2$trz)$_3$](D-CSA) in the LS and HS regimes (Fig. 3). The spectra show temperature-dependent line-shape variations across both the L$_3$ and L$_2$ edges, consistent with the established sensitivity of Fe L-edge XAS to changes in 3d occupancy, ligand-field strength, and multiplet structure in Fe(II) coordination environments [24, 35-37,39,40]. In an octahedral Fe(II) system, the SCO transition corresponds to a change from a low-spin configuration to a high-spin configuration, and this reconfiguration produces measurable spectral-weight redistribution between features associated with the different final-state manifolds accessed in the absorption process environments [24, 35-37,39,40]. The observed modifications to the L$_3$ region therefore, provide a direct spectroscopic indicator that the electronic configuration at the Fe site changes across the same temperature window identified by magnetometry. The XAS results, when considered alongside the CD data, support an interpretation in which the reduction of optical activity in the HS state is coupled to the SCO-driven reorganization of the Fe 3d-derived electronic structure and its hybridization with ligand states [44].

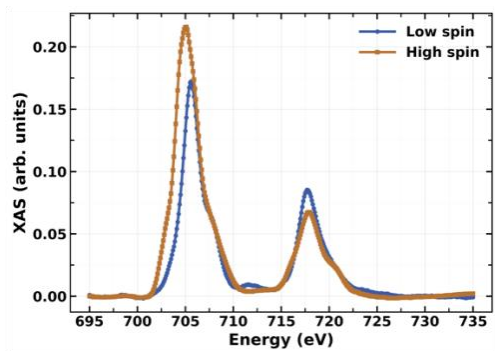

Fig 3. X-ray absorption spectra recorded at the Fe $L_{3,2}$ absorption edge in the low and high spin state for $[Fe(NH_2trz)_3](D\text{-}CSA)$.

*Electrical transport*

Electrical transport measurements were carried out to assess charge conduction in devices based on $[Fe(NH_2trz)_3](X\text{-}CSA)$, (where trz= triazole, CSA= camphorsulphonate and X= L/D) films deposited on Au interdigitated electrodes, and to evaluate how the electrical response behaves in the LS and HS regimes. Two-terminal current–voltage (I–V) characteristics exhibit pronounced hysteresis under bias sweeps in both spin states (Fig. 4a). Specifically, the measured current at a given applied voltage depends on the sweep direction and on the prior bias history, rather than collapsing onto a single-valued curve. Such hysteretic behavior is commonly associated with field-driven redistribution of mobile ionic species and interfacial polarization within dielectric or mixed ionic–electronic conductors, particularly in systems containing counterions that can migrate under an applied electric field [30-34]. In this framework, bias application can drive ions toward or away from electrodes, modify local band bending or injection barriers, and generate space-charge regions that persist on experimental timescales, all of which can produce memory-like behavior in I–V sweeps [30-34]. The observation that hysteresis is present in both LS and HS regimes indicates that these non-equilibrium processes remain operative across the temperature range studied and are not limited to a single spin state.

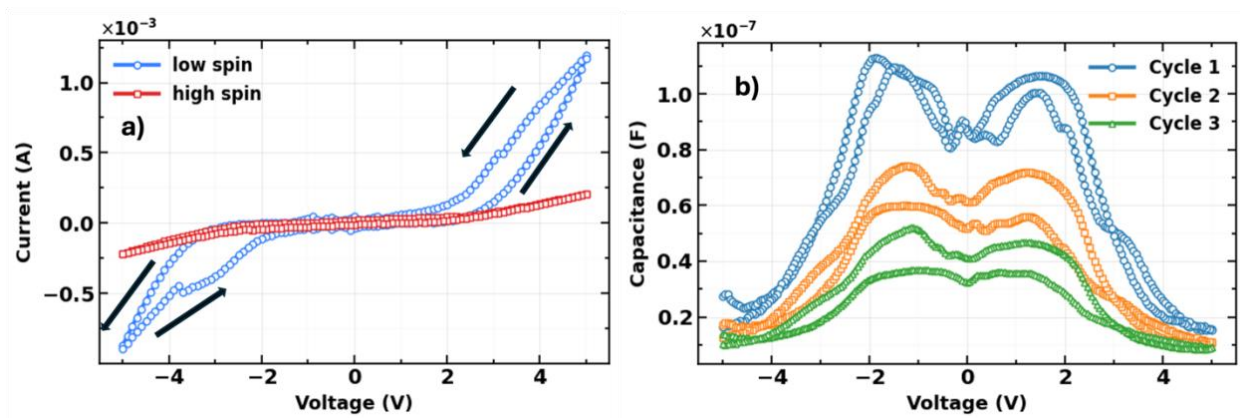

Fig 4. a) I(V) characteristics of a two terminal device fabricated by depositing a thin film of [Fe(NH$_2$trz)$_3$](D-CSA) on pre patterned Au-interdigitated electrodes. The measurements are recorded in the low spin state (blue) as well as the high spin state (red). The arrows indicate the voltage ramp direction. b) Small signal capacitance versus voltage measured at 1 kHz across multiple cycles for the same two terminal device.

Complementary capacitance–voltage (C–V) measurements at 1 kHz provide additional evidence for bias-conditioned polarization dynamics in the same devices (Fig. 4b). The capacitance response evolves over repeated voltage cycling, indicating that the dielectric properties sensed at small signal are not purely geometric or static, but instead depend on the prior electrical conditioning of the device. Cycling-dependent capacitance is consistent with gradual rearrangement of internal charge distributions, including redistribution of mobile ions, slow dipolar polarization, or the formation and relaxation of interfacial charge layers [30, 31]. At the measurement frequency employed, such processes can contribute to the effective capacitance if their characteristic timescales are comparable to, or slower than, the AC probing period and if the internal charge configuration is altered by the DC bias sweep. The combined presence of I–V hysteresis and cycling-dependent C–V behavior therefore supports a transport regime in which ionic migration and interfacial polarization play a dominant role under the measurement conditions [30-34].

Taken together, the magnetic, optical, X-ray spectroscopic, and electrical results provide a coherent picture of the temperature-driven switching behavior in [Fe(NH$_2$trz)$_3$](X-CSA), (where trz= triazole, CSA= camphorsulphonate and X= L/D. Magnetometry defines a cooperative SCO transition near room temperature for both enantiomers. CD spectroscopy shows that both compounds exhibit a larger chiroptical response in the LS state and a reduced response in the HS state, establishing a reproducible optical contrast linked to the spin state. Fe L-edge XAS directly evidences the accompanying Fe-centered electronic reconfiguration across the same LS and HS regimes. In devices incorporating these materials, transport is characterized by hysteresis and history-dependent capacitance consistent with an ionic-migration-dominated electrical response, indicating that the measured conduction and polarization are strongly influenced by mobile species and interfacial charging rather than purely electronic conduction.

**Conclusion**

We show that the enantiomeric pair in [Fe(NH$_2$trz)$_3$](X-CSA), (where trz= triazole, CSA= camphorsulphonate and X= L/D undergoes a cooperative, near-room-temperature spin crossover and exhibits a clear spin-state-dependent chiroptical response. The circular dichroism is strong in the low-spin state and strongly reduced in the high-spin state for both enantiomers, while Fe L$_{3,2}$-edge X-ray absorption confirms the corresponding Fe-centered electronic reconfiguration across the transition.

Although this system combines chirality with a thermally addressable spin transition, transport in devices based on it is dominated by ionic migration and interfacial polarization, as evidenced by pronounced I–V hysteresis and cycling-dependent C–V behavior. In such a regime, a stable, controllable CISS response cannot be reliably realized because the electrical characteristics are governed by time- and history-dependent ionic redistribution rather than electronic transmission. Consequently, despite having the key molecular ingredients for a switchable CISS effect, this material is not suitable for chiral spintronic device implementations under the present device configuration and measurement conditions.

**References**
[1] Brooker, S. Spin crossover with thermal hysteresis: practicalities and lessons learnt. Chemical Society Reviews 2015, 44 (10), 2880–2892. https://doi.org/10.1039/C4CS00376D.
[2] Nicolazzi, W.; Bousseksou, A. Thermodynamical aspects of the spin crossover phenomenon. Comptes Rendus Chimie 2018, 21 (12), 1060–1074. https://doi.org/10.1016/j.crci.2018.10.003.
[3] Stoleriu, L.; Chakraborty, P.; Hauser, A.; Stancu, A.; Enachescu, C. J. Thermal hysteresis in spin-crossover compounds studied within a mechanoelastic model. Physical Review B 2011, 84, 134102. https://doi.org/10.1103/PhysRevB.84.134102.
[4] Liu, W.; Bao, X.; Mao, L.-L.; Tucek, J.; Zboril, R.; Liu, J.-L.; Guo, F.-S.; Ni, Z.-P.; Tong, M.-L. A chiral spin crossover metal-organic framework. Chemical Communications 2014, 50, 4059-4061. https://doi.org/10.1039/C3CC48935C.
[5] Gural'skiy, I. A.; Reshetnikov, V. A.; Szebesczyk, A.; Gumienna-Kontecka, E.; Marynin, A. I.; Shylin, S. I.; Ksenofontov, V.; Fritsky, I. O. Chiral spin crossover nanoparticles and gels with switchable circular dichroism. Journal of Materials Chemistry C 2015, 3, 4737-4741. https://doi.org/10.1039/C5TC00161G.
[6] Sunatsuki, Y.; Ikuta, Y.; Matsumoto, N.; Ohta, H.; Kojima, M.; Iijima, S.; Hayami, S.; Maeda, Y.; Kaizaki, S.; Dahan, F.; Tuchagues, J.-P. An unprecedented homochiral mixed-valence spin-crossover compound. Angewandte Chemie International Edition 2003, 42, 1614-1618. https://doi.org/10.1002/anie.200250399.
[7] Sunatsuki, Y.; Ohta, H.; Kojima, M.; Ikuta, Y.; Goto, Y.; Matsumoto, N.; Iijima, S.; Akashi, H.; Kaizaki, S.; Dahan, F.; Tuchagues, J.-P. Supramolecular spin-crossover iron complexes based on imidazole-imidazolate hydrogen bonds. Inorganic Chemistry 2004, 43, 4154-4171. https://doi.org/10.1021/ic0498384.
[8] Sato, T.; Iijima, S.; Kojima, M.; Matsumoto, N. Assembling into chiral crystal of spin crossover iron(II) complex. Chemistry Letters 2009, 38, 178-179. https://doi.org/10.1246/cl.2009.178.
[9] Tian, L.; Pang, C.-Y.; Zhang, F.-L.; Qin, L.-F.; Gu, Z.-G.; Li, Z. Toward chiral iron(II) spin-crossover grafted resins with switchable circular dichroism. Inorganic Chemistry Communications 2015, 53, 55-59. https://doi.org/10.1016/j.inoche.2015.01.017.
[10] Kelly, C. T.; Jordan, R.; Felton, S.; Muller-Bunz, H.; Morgan, G. G. Spontaneous chiral resolution of a Mn(III) spin-crossover complex with high temperature 80 K hysteresis. Chemistry - A European Journal 2023, 29, e202300275. https://doi.org/10.1002/chem.202300275.

[11] Ru, J.; Yu, F.; Shi, P.-P.; Jiao, C.-Q.; Li, C.-H.; Xiong, R.-G.; Liu, T.; Kurmoo, M.; Zuo, J.-L. Three properties in one coordination complex: chirality, spin crossover, and dielectric switching. European Journal of Inorganic Chemistry 2017, 2017, 3144-3149. https://doi.org/10.1002/ejic.201700609.

[12] Suryadevara, N.; Pausch, A.; Moreno-Pineda, E.; Mizuno, A.; Burck, J.; Baksi, A.; Hochdorffer, T.; Salitros, I.; Ulrich, A. S.; Kappes, M. M.; Schunemann, V.; Klopper, W.; Ruben, M. Chiral resolution of spin-crossover active iron(II) [2x2] grid complexes. Chemistry - A European Journal 2021, 27, 15172-15180. https://doi.org/10.1002/chem.202101432.

[13] Kucheriv, O. I.; Oliynyk, V. V.; Zagorodnii, V. V.; Launets, V. L.; Fritsky, I. O.; Gural'skiy, I. A. New applications of spin-crossover complexes: microwave absorption, chirooptical switching and enantioselective detection. In Modern Magnetic and Spintronic Materials: Properties and Applications; Kaidatzis, A.; Sidorenko, S.; Vladymyrskyi, I.; Niarchos, D., Eds.; Springer: Dordrecht, 2020; pp 119-143. https://doi.org/10.1007/978-94-024-2034-0_6.

[14] Ren, D.-H.; Qiu, D.; Pang, C.-Y.; Li, Z.; Gu, Z.-G. Chiral tetrahedral iron(II) cages: diastereoselective subcomponent self-assembly, structure interconversion and spin-crossover properties. Chemical Communications 2015, 51, 788-791. https://doi.org/10.1039/C4CC08041F.

[15] Charytanowicz, T.; Dziedzic-Kocurek, K.; Kumar, K.; Ohkoshi, S.-I.; Chorazy, S.; Sieklucka, B. Chirality and spin crossover in iron(II)-octacyanidorhenate(V) coordination polymers induced by the pyridine-based ligand's positional isomer. Crystal Growth & Design 2023, 23, 4052-4064. https://doi.org/10.1021/acs.cgd.2c01462.

[16] Iazzolino, A.; Hamouda, A. O.; Naim, A.; Rosa, P.; Freysz, E. Nonlinear optical properties and application of a chiral spin crossover compound. In 2017 European Conference on Lasers and Electro-Optics and European Quantum Electronics Conference (CLEO/Europe-EQEC); IEEE, 2017; p 338.

[17] Qin, L.-F.; Pang, C.-Y.; Han, W.-K.; Zhang, F.-L.; Tian, L.; Gu, Z.-G.; Ren, X.; Li, Z. Optical recognition of alkyl nitrile by a homochiral iron(II) spin crossover host. CrystEngComm 2015, 17, 7956-7963. https://doi.org/10.1039/C5CE01617G.

[18] Zhao, X.-H.; Deng, Y.-F.; Huang, J.-Q.; Liu, M.; Zhang, Y.-Z. Solvated/desolvated homochiral Fe(II) complexes showing distinct bidirectional photo-switching due to a hidden state. Inorganic Chemistry Frontiers 2024, 11, 808-816. https://doi.org/10.1039/D3QI02167J.

[19] Ma, T.-T.; Sun, X.-P.; Yao, Z.-S.; Tao, J. Homochiral versus racemic polymorphs of spin-crossover iron(II) complexes with reversible LIESST effect. Inorganic Chemistry Frontiers 2020, 7, 1196-1204. https://doi.org/10.1039/C9QI01590F.

[20] Bartual-Murgui, C.; Pineiro-Lopez, L.; Valverde-Munoz, F. J.; Munoz, M. C.; Seredyuk, M.; Real, J. A. Chiral and racemic spin crossover polymorphs in a family of mononuclear iron(II) compounds. Inorganic Chemistry 2017, 56, 13535-13546. https://doi.org/10.1021/acs.inorgchem.7b02272.

[21] Regueiro, A.; Garcia-Lopez, V.; Forment-Aliaga, A.; Clemente-Leon, M. Chiral spin-crossover complexes based on an enantiopure Schiff base ligand with three chiral carbon centers. Dalton Transactions 2024, 53, 10637-10643. https://doi.org/10.1039/D4DT00924J.

[22] Ortega-Villar, N.; Munoz, M. C.; Real, J. A. Symmetry breaking in iron(II) spin-crossover molecular crystals. Magnetochemistry 2016, 2, 16. https://doi.org/10.3390/magnetochemistry2010016.

[23] Shatruk, M.; Phan, H.; Chrisostomo, B. A.; Suleimenova, A. Symmetry-breaking structural phase transitions in spin crossover complexes. Coordination Chemistry Reviews 2015, 289-290, 62-73. https://doi.org/10.1016/j.ccr.2014.09.018.

[24] Ekanayaka, T. K.; Ungor, O.; Hu, Y.; Mishra, E.; Phillips, J. P.; Dale, A. S.; Yazdani, S.; Wang, P.; McElveen, K. A.; Zaz, M. Z.; Zhang, J.; N'Diaye, A. T.; Klewe, C.; Shafer, P.; Lai, R. Y.; Streubel, R.; Cheng, R.; Shatruk, M.; Dowben, P. A. Perturbing the spin state and conduction of Fe(II) spin crossover complexes with TCNQ. Materials Chemistry and Physics 2023, 296, 127276. https://doi.org/10.1016/j.matchemphys.2022.127276.

[25] Peacock, R. D.; Stewart, B. Natural circular dichroism in X-ray spectroscopy. The Journal of Physical Chemistry B 2001, 105, 351-360. https://doi.org/10.1021/jp001946y.

[26] Gural'skiy, I. A.; Reshetnikov, V. A.; Szebesczyk, A.; Gumienna-Kontecka, E.; Marynin, A. I.; Shylin, S. I.; Ksenofontov, V.; Fritsky, I. O. Chiral spin crossover nanoparticles and gels with switchable circular dichroism. J. Mater. Chem. C 2015, 3 (18), 4737–4741. https://doi.org/10.1039/C5TC00161G.

[27] Naaman, R.; Waldeck, D. H. Chiral-Induced Spin Selectivity Effect. J. Phys. Chem. Lett. 2012, 3 (16), 2178–2187. https://doi.org/10.1021/jz300793y.

[28] Bloom, B. P.; Paltiel, Y.; Naaman, R.; Waldeck, D. H. Chiral Induced Spin Selectivity. Chem. Rev. 2024, 124 (4), 1950–1991. https://doi.org/10.1021/acs.chemrev.3c00661.

[29] Kahn, O.; Martinez, C. J. Spin-Transition Polymers: From Molecular Materials Toward Memory Devices. Science 1998, 279 (5347), 44–48. https://doi.org/10.1126/science.279.5347.44.

[30] Egginger, M.; Irimia-Vladu, M.; Schwödiauer, R.; Tanda, A.; Frischauf, I.; Bauer, S.; Sariciftci, N. S. Mobile Ionic Impurities in Poly(vinyl alcohol) Gate Dielectric: Possible Source of the Hysteresis in Organic Field-Effect Transistors. Adv. Mater. 2008, 20 (5), 1018–1022. https://doi.org/10.1002/adma.200701479.

[31] Waser, R.; Aono, M. Nanoionics-based resistive switching memories. Nat. Mater. 2007, 6 (11), 833–840. https://doi.org/10.1038/nmat2023.

[32] Jeong, S.; Han, C.-H.; Kwak, B.; Koo, R.-H.; Cho, Y.; Kim, J.; Lee, J.-H.; Kwon, D.; Shin, W. Unraveling ionic switching dynamics in high-k dielectric double-gate transistors via low-frequency noise spectroscopy. Nano Converg. 2025, 12 (1), 48. https://doi.org/10.1186/s40580-025-00512-2.

[33] Jonscher, A. K. Dielectric Relaxation in Solids; Chelsea Dielectrics Press: London, 1983; ISBN 0-9508711-0-9.

[34] Barsoukov, E.; Macdonald, J. R., Eds. Impedance Spectroscopy: Theory, Experiment, and Applications, 2nd ed.; Wiley-Interscience: Hoboken, NJ, 2005; https://doi.org/10.1002/0471716243; ISBN 978-0-471-64749-2

[35] Phillips, J. P.; Yazdani, S.; Soruco, J.; Oles, J.; Ekanayaka, T. K.; Mishra, E.; Wang, P.; Zaz, M. Z.; Liu, J.; N'Diaye, A. T.; Shatruk, M.; Dowben, P. A.; Cheng, R. Conductance

fluctuations in cobalt valence tautomer molecular thin films. Dalton Trans. 2024, 53, 17571-17580. https://doi.org/10.1039/D4DT02213K.

[36] Mishra, E.; Chin, W.; McElveen, K. A.; Ekanayaka, T. K.; Zaz, M. Z.; Viswan, G.; Zielinski, R.; N'Diaye, A. T.; Shapiro, D.; Lai, R. Y.; Streubel, R.; Dowben, P. A. Electronic transport properties of spin-crossover polymer plus polyaniline composites with Fe3O4 nanoparticles. J. Phys. Mater. 2024, 7(1), 015010. https://doi.org/10.1088/2515-7639/ad1b35.

[37] Dale, A. S.; Yazdani, S.; Ekanayaka, T. K.; Mishra, E.; Hu, Y.; Dowben, P. A.; Freeland, J. W.; Zhang, J.; Cheng, R. Direct observation of the magnetic anisotropy of an Fe(II) spin crossover molecular thin film. J. Phys. Mater. 2023, 6(3), 035010. https://doi.org/10.1088/2515-7639/ace21a.

[38] Mishra, E.; Ekanayaka, T. K.; Panagiotakopoulos, T.; Le, D.; Rahman, T. S.; Wang, P.; McElveen, K. A.; Phillips, J. P.; Zaz, M. Z.; Yazdani, S.; N'Diaye, A. T.; Lai, R. Y.; Streubel, R.; Cheng, R.; Shatruk, M.; Dowben, P. A. Electronic structure of cobalt valence tautomeric molecules in different environments. Nanoscale 2023, 15, 2044-2053. https://doi.org/10.1039/D2NR06834F.

[39] Ekanayaka, T. K.; Kurz, H.; McElveen, K. A.; Hao, G.; Mishra, E.; N'Diaye, A. T.; Lai, R. Y.; Weber, B.; Dowben, P. A. Evidence for surface effects on the intermolecular interactions in Fe(II) spin crossover coordination polymers. Phys. Chem. Chem. Phys. 2022, 24, 883-894. https://doi.org/10.1039/D1CP04243B.

[40] Ekanayaka, T. K.; Kurz, H.; Dale, A. S.; Hao, G.; Mosey, A.; Mishra, E.; N'Diaye, A. T.; Cheng, R.; Weber, B.; Dowben, P. A. Probing the unpaired Fe spins across the spin crossover of a coordination polymer. Mater. Adv. 2021, 2, 760-768. https://doi.org/10.1039/D0MA00612B.

[41] Sugahara, A.; Kamebuchi, H.; Okazawa, A.; Enomoto, M.; Kojima, N. Control of Spin-Crossover Phenomena in One-Dimensional Triazole-Coordinated Iron(II) Complexes by Means of Functional Counter Ions. Inorganics 2017, 5 (3), 50. https://doi.org/10.3390/inorganics5030050.

[42] Garcia, Y. Crystal Engineering of FeII Spin Crossover Coordination Polymers Derived from Triazole or Tetrazole Ligands. CHIMIA 2013, 67 (6), 411–418. https://doi.org/10.2533/chimia.2013.411.

[43] Gütlich, P.; Goodwin, H. A., Eds. Spin Crossover in Transition Metal Compounds I–III; Topics in Current Chemistry; Springer: Berlin, 2004; Vols. 233–235.

[44] de Groot, F. M. F.; Hu, Z. W.; Lopez, M. F.; Kaindl, G.; Guillot, F.; Tronc, M. Differences between L3 and L2 X-ray absorption spectra of transition metal compounds. J. Chem. Phys. 1994, 101 (8), 6570–6576.